\documentclass[preprint,12pt]{elsarticle}
\usepackage{graphicx}
\usepackage{amssymb}
\usepackage{amsmath}
\usepackage{amsfonts}
\usepackage{slashed} 
\usepackage[dvipsnames]{xcolor}

\newcommand{\bs}[1]{\boldsymbol{#1}} 
\newcommand{\nopieft}{\mbox{$\slashed{\pi}$EFT}} 
 
\newcommand{\Lag}{{\cal L}} 
\newcommand{\be}{\begin{equation}} 
\newcommand{\ee}{\end{equation}} 
\newcommand{\rvec}{{\bs{r}}}

\journal{Physics Letters B} 

\begin{document}

\begin{frontmatter} 

\title{The onset of $\Lambda\Lambda$ hypernuclear binding} 
\author[a]{L.~Contessi} 
\address[a]{Racah Institute of Physics, The Hebrew University, 
Jerusalem 91904, Israel}
\author[b,c]{M.~Sch{\"a}fer} 
\address[b]{Czech Technical University in Prague, Faculty of Nuclear Sciences 
and Physical Engineering, B\v{r}ehov\'{a} 7, 11519 Prague 1, Czech Republic} 
\address[c]{Nuclear Physics Institute, Academy of Sciences of Czech Republic, 
250 69 \v{R}e\v{z}, Czech Republic} 
\author[a]{N.~Barnea} 
\author[a]{A.~Gal\corref{cor1}~} 
\cortext[cor1]{corresponding author: Avraham Gal, avragal@savion.huji.ac.il}
\author[c]{J.~Mare\v{s}}
\date{\today}

\begin{abstract} 

Binding energies of light, $A\leq 6$, $\Lambda\Lambda$ hypernuclei are 
calculated using the stochastic variational method in a pionless effective 
field theory ({\nopieft}) approach at leading order with the purpose of 
assessing critically the onset of binding in the strangeness ${\cal S}=-2$ 
hadronic sector. The {\nopieft} input in this sector consists of (i) a 
$\Lambda\Lambda$ contact term constrained by the $\Lambda\Lambda$ scattering 
length $a_{\Lambda\Lambda}$, using a range of values compatible with 
$\Lambda\Lambda$ correlations observed in relativistic heavy ion collisions, 
and (ii) a $\Lambda\Lambda N$ contact term constrained by the only available 
$A\leq 6$ $\Lambda\Lambda$ hypernucler binding energy datum of $^{~~6}_{
\Lambda\Lambda}$He. The recently debated neutral three-body and four-body 
systems $^{~~3}_{\Lambda\Lambda}$n and $^{~~4}_{\Lambda\Lambda}$n are found 
unbound by a wide margin. A relatively large value of $|a_{\Lambda\Lambda}|
\gtrsim 1.5$ fm is needed to bind $^{~~4}_{\Lambda\Lambda}$H, thereby 
questioning its particle stability. In contrast, the particle stability of
the $A=5$ $\Lambda\Lambda$ hypernuclear isodoublet $^{~~5}_{\Lambda\Lambda}
$H--$^{~~5}_{\Lambda\Lambda}$He is robust, with $\Lambda$ separation energy 
of order 1~MeV. 

\end{abstract}


\end{frontmatter}

\section{Introduction}
\label{sec:intro}

Single-$\Lambda$ and double-$\Lambda$ ($\Lambda\Lambda$) 
hypernuclei provide a unique extension of nuclear physics into strange 
hadronic matter~\cite{SBG00}. Whereas the behavior of a single $\Lambda$ 
hyperon in atomic nuclei has been deduced quantitatively by studying 
$\Lambda$ hypernuclei ($_{\Lambda}^{\rm A}$Z) from $A$=3 to 208~\cite{GHM16}, 
only three $\Lambda\Lambda$ hypernuclei ($^{~\rm A}_{\Lambda\Lambda}$Z) are 
firmly established: the lightest known $^{~~6}_{\Lambda\Lambda}$He Nagara 
event~\cite{LL6He} and two heavier ones, $^{10}_{\Lambda\Lambda}$Be and 
$^{13}_{\Lambda\Lambda}$B~\cite{HN18}. Remarkably, their binding energies 
come out consistently in shell-model calculations~\cite{GM11}. 
Few ambiguous emulsion events from KEK~\cite{KEK09} and J-PARC~\cite{E0719} 
have also been reported. However, and perhaps more significant is the absence 
of any good data on the onset of $\Lambda\Lambda$ hypernuclear binding for 
$A<6$. In distinction from the heavier species, these very light $s$-shell 
species, if bound, could be more affected by microscopic strangeness 
${\cal S}=-2$ dynamics. An obvious issue is the effect of a possible $\Xi N$ 
dominated $H$ dibaryon resonance some 20--30~MeV above the $\Lambda\Lambda$ 
threshold~\cite{halqcd12,HM12} on $\Lambda\Lambda$ hypernuclear binding in 
general. 

Several calculations of light $A<6$ $s$-shell $\Lambda\Lambda$ hypernuclei 
using $\Lambda\Lambda$ interactions fitted to $^{~~6}_{\Lambda\Lambda}$He 
suggest a fairly weak $\Lambda\Lambda$ interaction, with the onset of 
$\Lambda\Lambda$ hypernuclear binding defered to $A=4$. Indeed, a slightly 
bound $I=0$ $^{~~4}_{\Lambda\Lambda}$H(1$^+$) was found in $\Lambda\Lambda 
pn$ four-body calculations by Nemura et al.~\cite{NAM03,NSAM05} but not 
in a four-body calculation by Filikhin and Gal~\cite{FG02a} who nonetheless 
got it bound as a $\Lambda\Lambda d$ cluster. Unfortunately, the AGS-E906 
counter experiment \cite{Ahn01} searching for light $\Lambda\Lambda$ 
hypernuclei failed to provide conclusive evidence for the particle stability 
of $^{~~4}_{\Lambda\Lambda}$H~\cite{RH07,poch19}. Interestingly, the neutral 
four-body system $^{~~4}_{\Lambda\Lambda}$n has been assigned in 
Ref.~\cite{poch19} to the main yet unexplained signal observed by AGS-E906. 
Recent few-body calculations of $^{~~4}_{\Lambda\Lambda}$n~\cite{RWZ15,GVV17} 
diverge on its particle stability, but since none was constrained by the 
$^{~~6}_{\Lambda\Lambda}$He binding energy datum, no firm conclusion can be 
drawn yet. 

In the present work we study the light $A\leq 6$ $s$-shell $\Lambda\Lambda$ 
hypernuclei together with their nuclear and $\Lambda$ hypernuclear cores 
at leading-order (LO) {\nopieft}. The {\nopieft} approach was first applied 
to few-nucleon atomic nuclei in Refs.~\cite{Kol99,BHK00} and recently also 
in lattice calculations of nuclei \cite{BCG15,KBG15,CLP17,KPDB17} and to 
$s$-shell single-$\Lambda$ hypernuclei~\cite{CBG18}. Focusing on {\nopieft} 
applications to ${\cal S}=-2$ light systems, we note $\Lambda$-$\Lambda$-core 
LO calculations done for $A=4$~\cite{Ando14a} and separately for $A=6$~\cite{
Ando14b}, which therefore limits their predictive power. Among past non-EFT 
studies, the only work that covers {\it all} $s$-shell $\Lambda\Lambda$ 
hypernuclei is by Nemura et al.~\cite{NSAM05} who used simulated forms of 
outdated hard-core $YN$ and $YY$ Nijmegen potentials~\cite{NimDF}. No chiral 
EFT ($\chi$EFT) calculations of $\Lambda\Lambda$ hypernuclei have been 
reported, although $\chi$EFT representations of the $\Lambda\Lambda$ 
interaction at LO~\cite{PHM07} and NLO~\cite{HMP16} do exist. Hence, the 
present LO {\nopieft} work is the first comprehensive EFT application to light 
$\Lambda\Lambda$ hypernuclei, and could be generalized in principle to study 
multi-$\Lambda$ hypernuclei and strange hadronic matter.  

The {\nopieft} baryonic Lagrangian governing multi-$\Lambda$ hypernuclei 
requires, at LO, one $\Lambda\Lambda$ and one $\Lambda\Lambda N$ interaction 
terms beyond the interaction terms involved in the description of 
single-$\Lambda$ hypernuclei. We fit the new $\Lambda\Lambda$ contact term 
to a $\Lambda\Lambda$ scattering length value spanning a range of values, 
$-0.5$~fm to $-1.9$~fm, suggested by analyses of the $\Lambda\Lambda$ 
invariant mass spectrum~\cite{GHH12} measured in the $^{12}$C$(K^-,K^+)$ 
reaction at the KEK-PS~\cite{Yoon07} and of $\Lambda\Lambda$ correlations 
\cite{MFO15} extracted from ultra-relativistic Au+Au collisions at the 
RHIC-STAR experiment~\cite{STAR15}. The corresponding $\Lambda\Lambda$ 
interactions are weakly atractive, far from producing a $\Lambda\Lambda$ 
bound state. Recent analyses of LHC-ALICE experiments reach similar 
conclusions, but leave room also for a $\Lambda\Lambda$ bound 
state~\cite{ALICE19}. For each choice of $\Lambda\Lambda$ contact term we 
determine a $\Lambda\Lambda N$ three-body contact term, promoted to LO, by 
fitting to $\Delta B_{\Lambda\Lambda}(^{~~6}_{\Lambda\Lambda}$He)=$B_{\Lambda
\Lambda}(^{~~6}_{\Lambda\Lambda}$He)$-2B_{\Lambda}(_{\Lambda}^5$He)=0.67$\pm
$0.17~MeV. With such $\Lambda\Lambda$ and $\Lambda\Lambda N$ contact-term 
input our {\nopieft} scheme exhibits renormalization scale invariance in the 
limit of point-like interactions, demonstrating that no $N\geq 4$ $N$-body 
contact term is required at LO, as first shown by Platter et al.~\cite{PHM05} 
for a four-nucleon system. Applying this scheme to explore $A=3,4,5$ $\Lambda
\Lambda$ hypernuclei, we find that unless $|a_{\Lambda\Lambda}|\gtrsim 
1.5$~fm, $^{~~4}_{\Lambda\Lambda}$H is unlikely to be particle stable. 
The neutral systems $^{~~3}_{\Lambda\Lambda}$n and $^{~~4}_{\Lambda\Lambda}$n 
are found unstable by a wide margin. A robust particle stability is 
established for the $^{~~5}_{\Lambda\Lambda}$H--$^{~~5}_{\Lambda\Lambda}$He 
$A=5$ isodoublet, with $\Lambda$ separation energy of order 1~MeV, providing 
further support for a recent J-PARC proposal P75~\cite{P75} to produce 
$^{~~5}_{\Lambda\Lambda}$H. Possible extensions of our work are briefly 
discussed in the concluding section.

\section{Application of {\nopieft} to $\Lambda\Lambda$ hypernuclei} 
\label{sec:ext} 

With $\Lambda\Lambda$ one-pion exchange forbidden by isospin invariance, 
the lowest mass pseudoscalar meson exchange is provided by a short range 
$\eta$ exchange ($\approx$0.4~fm) which is rather weak in SU(3) flavor. 
Pions appear in the $\Lambda\Lambda$ dynamics through excitation to fairly 
high-lying $\Sigma\Sigma$ intermediate states. Therefore, a reasonable choice 
of a {\nopieft} breakup scale is 2$m_{\pi}$, same as argued for in our recent 
work on $\Lambda$ hypernuclei~\cite{CBG18}. Excitation from $\Lambda\Lambda$ 
states to the considerably lower mass $\Xi N$ intermediate states requires 
a shorter range $K$ meson exchange which, together with other short-range 
exchanges, is accounted for implicitly by the chosen {\nopieft} contact 
interactions. To provide a meaningful {\nopieft} expansion parameter we note 
that since $\Delta B_{\Lambda\Lambda}(^{~~6}_{\Lambda\Lambda}$He) is less 
than 1~MeV, considerably smaller than $B_\Lambda(^5_\Lambda$He), a $\Lambda$ 
momentum scale $Q$ in $^{~~6}_{\Lambda\Lambda}$He may be approximated by that 
in $^5_\Lambda$He~\cite{CBG18}, namely $p_{\Lambda}\approx\sqrt{2M_{\Lambda}
B_{\Lambda}}=83$~MeV/c, yielding a {\nopieft} expansion parameter $(Q/2m_{\pi}
)\approx 0.3$ and LO accuracy of order $(Q/2m_{\pi})^2\approx 0.09$. 

\begin{figure}[htb] 
\begin{center} 
\includegraphics[width=0.6\textwidth]{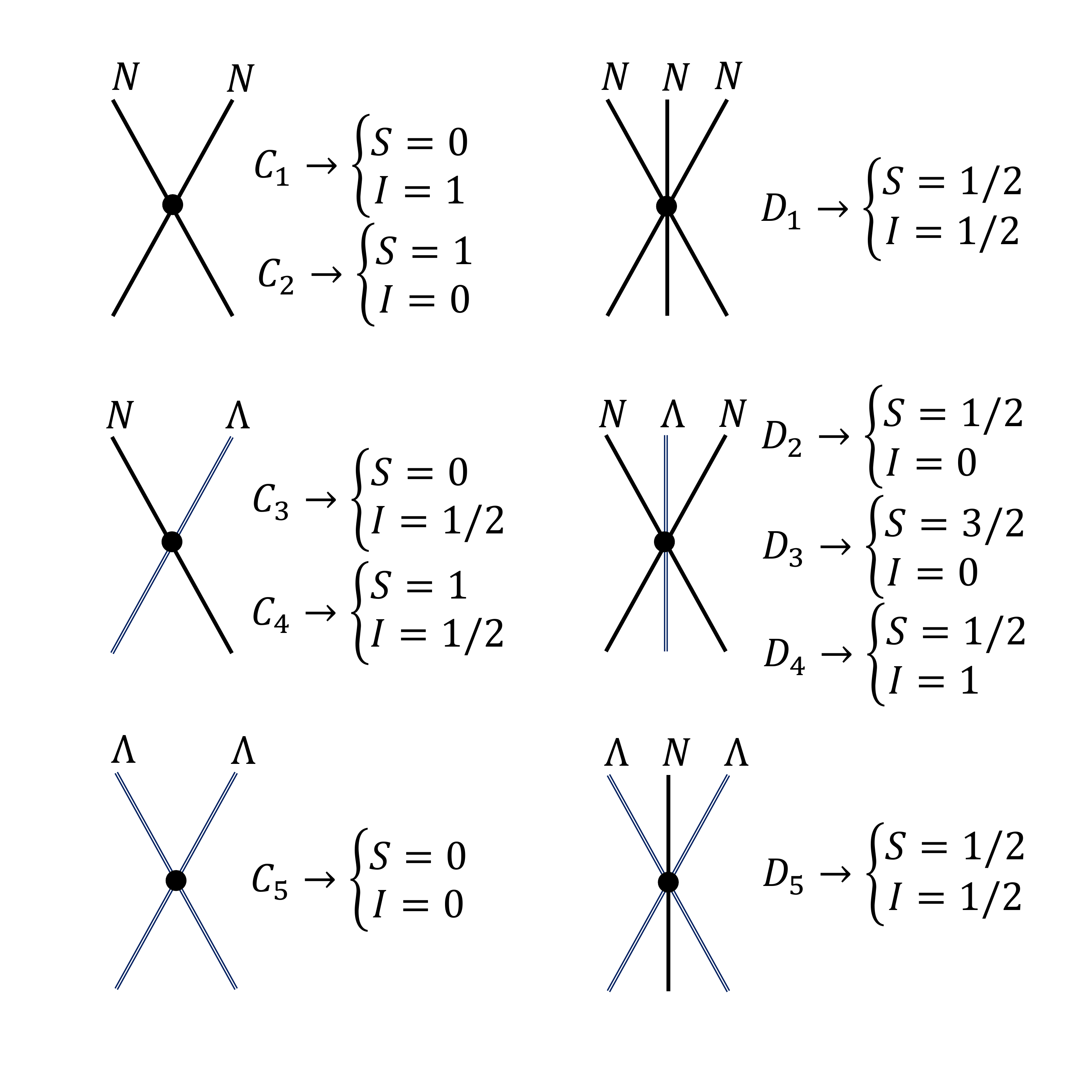} 
\caption{Diagrammatic presentation of two-body (left) and three-body 
(right) contact terms, and their associated LEC input ($C_1,\ldots,C_5$) 
and ($D_1,\ldots,D_5$) to a LO {\nopieft} calculation of light nuclei 
(upper) $\Lambda$ hypernuclei (middle) and $\Lambda\Lambda$ hypernuclei 
(lower), with values of spin $S$ and isospin $I$ corresponding to $s$-wave 
configurations.} 
\label{fig:diagrams} 
\end{center} 
\end{figure} 

To construct the appropriate {\nopieft} Lagrangian density at LO we 
follow our previous work on single-$\Lambda$ hypernuclei~\cite{CBG18}: 
\begin{equation} 
\Lag^{(\rm LO)} = \sum_B{B^\dagger (i\partial_0+\frac{\nabla^2}{2 M_B})B} 
  - {\cal V}_2 - {\cal V}_3, 
\label{eq:Lag} 
\end{equation}
where $B=(N,\Lambda)$ and ${\cal V}_2,{\cal V}_3$ consist of two-body and 
three-body $s$-wave contact interaction terms, each of which is associated 
with its own low-energy constant (LEC). These contact terms are shown 
diagrammatically in Fig.~\ref{fig:diagrams} and the corresponding LECs are 
listed alongside. Going from single-$\Lambda$ hypernuclei to multi-$\Lambda$ 
hypernuclei brings in one new $\Lambda\Lambda$ two-body LEC, $C_5$, and one 
new $\Lambda\Lambda N$ three-body LEC, $D_5$, each one labelled by the total 
Pauli-spin and isospin involved. This completes the set of LECs required to 
describe single-, double- and in general multi-$\Lambda$ hypernuclei at LO. 
Further contact terms, such as a three-body $\Lambda\Lambda\Lambda$ term, 
appear only at subleading orders.  

Following the procedure applied in Ref.~\cite{Kol99}, the two-body contact 
interaction term ${\cal V}_2$ gives rise to a two-body potential  
\begin{equation} 
V_2 = \sum_{IS}\,C_{\lambda}^{IS} \sum_{i<j} {\cal P}_{IS}(ij)
          \delta_\lambda(\rvec_{ij}), 
\label{eq:V2} 
\end{equation}
where ${\cal P}_{IS}$ are projection operators on $s$-wave $NN,\Lambda N,
\Lambda\Lambda$ pairs with isospin $I$ and spin $S$ values associated in 
Fig.~\ref{fig:diagrams} with two-body LECs. These LECs are fitted to 
low-energy two-body observables, e.g., to the corresponding $NN,\Lambda N,
\Lambda\Lambda$ scattering lengths. The subscript $\lambda$ attached to 
$C^{IS}$ in Eq.~(\ref{eq:V2}) stands for a momentum cutoff introduced in 
a Gaussian form to regularize the zero-range contact terms: 
\begin{equation}
\delta_\lambda(\rvec)=\left(\frac{\lambda}{2\sqrt{\pi}}\right)^3\,
\exp \left(-{\frac{\lambda^2}{4}}\rvec^2\right),  
\label{eq:gaussian} 
\end{equation}
thereby smearing a zero-range (in the limit $\lambda\to\infty$) Dirac 
$\delta^{(3)}(\rvec)$ contact term over distances~$\sim\lambda^{-1}$. The 
cutoff parameter $\lambda$ may be viewed as a scale parameter with respect 
to typical values of momenta $Q$. To make observables cutoff independent, 
the LECs must be properly renormalized. Truncating 
{\nopieft} at LO and using values of $\lambda$ higher than the breakup scale 
of the theory (here $\approx$2$m_{\pi}$), observables acquire a residual 
dependence $O(Q/\lambda)$ which diminishes with increasing $\lambda$. 

The three-body contact interaction, promoted to LO, gives rise to a three-body 
potential  
\begin{equation} 
V_3 = \sum_{\alpha IS}D_{\alpha\lambda}^{IS} \sum_{i<j<k}
{\cal Q}_{IS}(ijk)\left(\sum_{\rm cyc}\delta_\lambda(\rvec_{ij})
\delta_\lambda(\rvec_{jk})\right), 
\label{eq:V3} 
\end{equation} 
where ${\cal Q}_{IS}$ projects on $NNN$, $NN\Lambda$ and $\Lambda\Lambda 
N$ $s$-wave triplets with isospin $I$ and spin $S$ values associated in 
Fig.~\ref{fig:diagrams} with three-body LECs which are fitted to given binding 
energies. The subscript $\alpha$ distinguishes between the two $IS=\frac{1}{2}
\frac{1}{2}$ $NNN$ and $\Lambda\Lambda N$ triplets marked in the figure.

\begin{figure}[t!] 
\begin{center}
\includegraphics[width=0.7\textwidth]{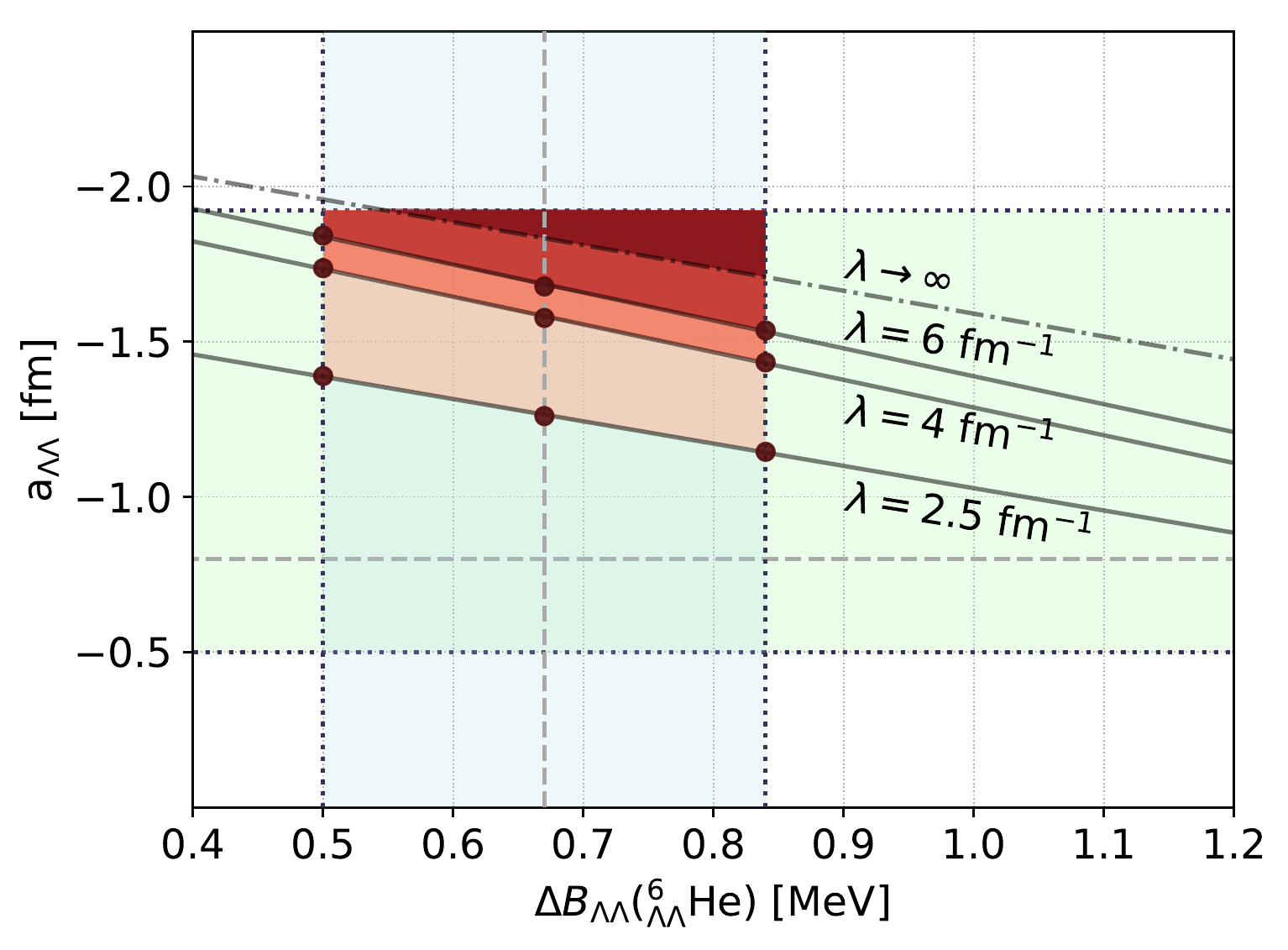} 
\caption{Minimum values of $|a_{\Lambda\Lambda}|$ for which $^{~~4}_{\Lambda
\Lambda}$H becomes bound are plotted, for several values of cutoff $\lambda$, 
as a function of $\Delta B_{\Lambda\Lambda}(^{~~6}_{\Lambda\Lambda}$He) using 
the Alexander[B] $\Lambda N$ interaction model~\cite{CBG18}. The vertical 
dotted lines mark the experimental uncertainty of $\Delta B_{\Lambda\Lambda}$. 
Horizontal lines mark the range of $a_{\Lambda\Lambda}$ values [$-0.5,\,-1.9$] 
fm suggested by studies of $\Lambda\Lambda$ correlations~\cite{GHH12,MFO15}, 
with a representative value of $a_{\Lambda\Lambda}=-0.8$~fm marked by a dashed 
line. The $\lambda\to\infty$ limit is reached assuming a $Q/\lambda$ 
asymptotic behavior, similar to the discussion around Eq.~(5) below.} 
\label{fig:LL4H} 
\end{center} 
\end{figure} 

Using two-body $V_2$ and three-body $V_3$ regularized contact interaction 
terms as described above, we solved the $A$-body Schr\"{o}dinger equation 
variationally by expanding the wave function $\Psi$ in a correlated 
Gaussian basis using the SVM. For a comprehensive review of this method, 
see Ref.~\cite{SVa98}. For a specific calculation of the three-body 
interaction matrix elements, see Ref.~\cite{Bazak16}.

\section{Results and discussion} 
\label{sec:res} 

We first discuss the case of $^{~~4}_{\Lambda\Lambda}$H, with $I=0$ and 
$J^\pi=1^+$, which following the brief discussion in the Introduction could 
signal the onset of $\Lambda\Lambda$ hypernuclear binding. For each of several 
given cutoff values $\lambda$ we searched for minimum values of $|a_{\Lambda
\Lambda}|$, as a function of $\Delta B_{\Lambda\Lambda}(^{~~6}_{\Lambda\Lambda}
$He), that would make $^{~~4}_{\Lambda\Lambda}$H particle stable. The choice 
of a specific value for this $\Delta B_{\Lambda\Lambda}$ determines the 
$\Lambda\Lambda N$ LEC necessary for the $^{~~4}_{\Lambda\Lambda}$H 
calculation, in addition to the $\Lambda\Lambda$ LEC determined by 
$a_{\Lambda\Lambda}$. The resulting values of $|a_{\Lambda\Lambda}|$ 
above which $^{~~4}_{\Lambda\Lambda}$H is particle stable are plotted in 
Fig.~\ref{fig:LL4H} as a function of $\Delta B_{\Lambda\Lambda}(^{~~6}_{
\Lambda\Lambda}$He). Choosing sufficiently large values of the cutoff 
$\lambda$, say $\lambda\gtrsim$~4~fm$^{-1}$, for which convergence to the 
renormalization scale invariance limit $\lambda\to\infty$ is seen explicitly 
in the figure, one concludes that $|a_{\Lambda\Lambda}|$ needs to be larger 
than $\approx$1.5~fm to bind $^{~~4}_{\Lambda\Lambda}$H. A $\Lambda\Lambda$ 
scattering length of such size would make the $\Lambda\Lambda$ interaction 
almost as strong as the $\Lambda N$ interaction, whereas most theoretical 
constructions, e.g. recent Nijmegen models, suggest that it is considerably 
weaker, say $|a_{\Lambda\Lambda}|\approx 0.8$~fm~\cite{Nim10}. For this reason 
we argue that $^{~~4}_{\Lambda\Lambda}$H is unlikely to be particle stable. 

\begin{table}[htb] 
\begin{center} 
\caption{$\Lambda$ separation energies $B_\Lambda(^{~~A}_{\Lambda\Lambda}$Z)
for $A$=3--6, calculated using $a_{\Lambda\Lambda}$=$-0.8$~fm, cutoff 
$\lambda$=4~fm$^{-1}$ and the Alexander[B] $\Lambda N$ interaction
model~\cite{CBG18}. In each row a $\Lambda\Lambda N$ LEC was fitted to the
underlined binding energy constraint.} 
\begin{tabular}{crrrrr}
\hline 
Constraint (MeV) & $^{~~3}_{\Lambda\Lambda}$n & $^{~~4}_{\Lambda\Lambda}$n &
$^{~~4}_{\Lambda\Lambda}$H & $^{~~5}_{\Lambda\Lambda}$H & $^{~~6}_{\Lambda
\Lambda}$He \\
\hline 
$\Delta B_{\Lambda\Lambda}(^{~~6}_{\Lambda\Lambda}$He)=$\underline{0.67}$ &
-- & -- & -- & 1.21 & 3.28 \\
$B_{\Lambda}(^{~~4}_{\Lambda\Lambda}$H)=$\underline{0.05}$ & -- & -- & 0.05 &
2.28 & 4.76 \\
$B(^{~~4}_{\Lambda\Lambda}$n)=$\underline{0.10}$ & -- & 0.10 & 0.86 & 4.89 &
7.89 \\
$B(^{~~3}_{\Lambda\Lambda}$n)=$\underline{0.10}$ & 0.10 & 15.15 & 18.40 &
22.13 & 25.66 \\
\hline 
\end{tabular} 
\label{tab:BLL} 
\end{center} 
\end{table} 

Using representative values $a_{\Lambda\Lambda}=-0.8$~fm and cutoff $\lambda
$=4~fm$^{-1}$, values for which according to Fig.~\ref{fig:LL4H} $^{~~4}_{
\Lambda\Lambda}$H is particle unstable, one may reduce the repulsive $\Lambda
\Lambda N$ LEC in order to make it particle stable. According to the first two 
rows in Table~\ref{tab:BLL}, this will overbind $^{~~6}_{\Lambda\Lambda}$He by 
$\approx$1.5~MeV. Reducing further the $\Lambda\Lambda N$ LEC one binds 
the neutral systems, first $^{~~4}_{\Lambda\Lambda}$n (third row) and then 
$^{~~3}_{\Lambda\Lambda}$n (fourth row), at a price of overbinding further 
$^{~~6}_{\Lambda\Lambda}$He. In fact, the particle stability of these 
$A=3,4$ neutral $\Lambda\Lambda$ systems is incompatible with the 
$^{~~6}_{\Lambda\Lambda}$He Nagara event binding energy datum for all values 
of cutoff $\lambda$ and scattering length $a_{\Lambda\Lambda}$ tested in 
Fig.~\ref{fig:LL4H}. These results suggest quantitatively that the $A=3,4$ 
light neutral $\Lambda\Lambda$ hypernuclei are unbound within a large margin. 

\begin{figure}[htb] 
\begin{center} 
\includegraphics[width=0.7\textwidth]{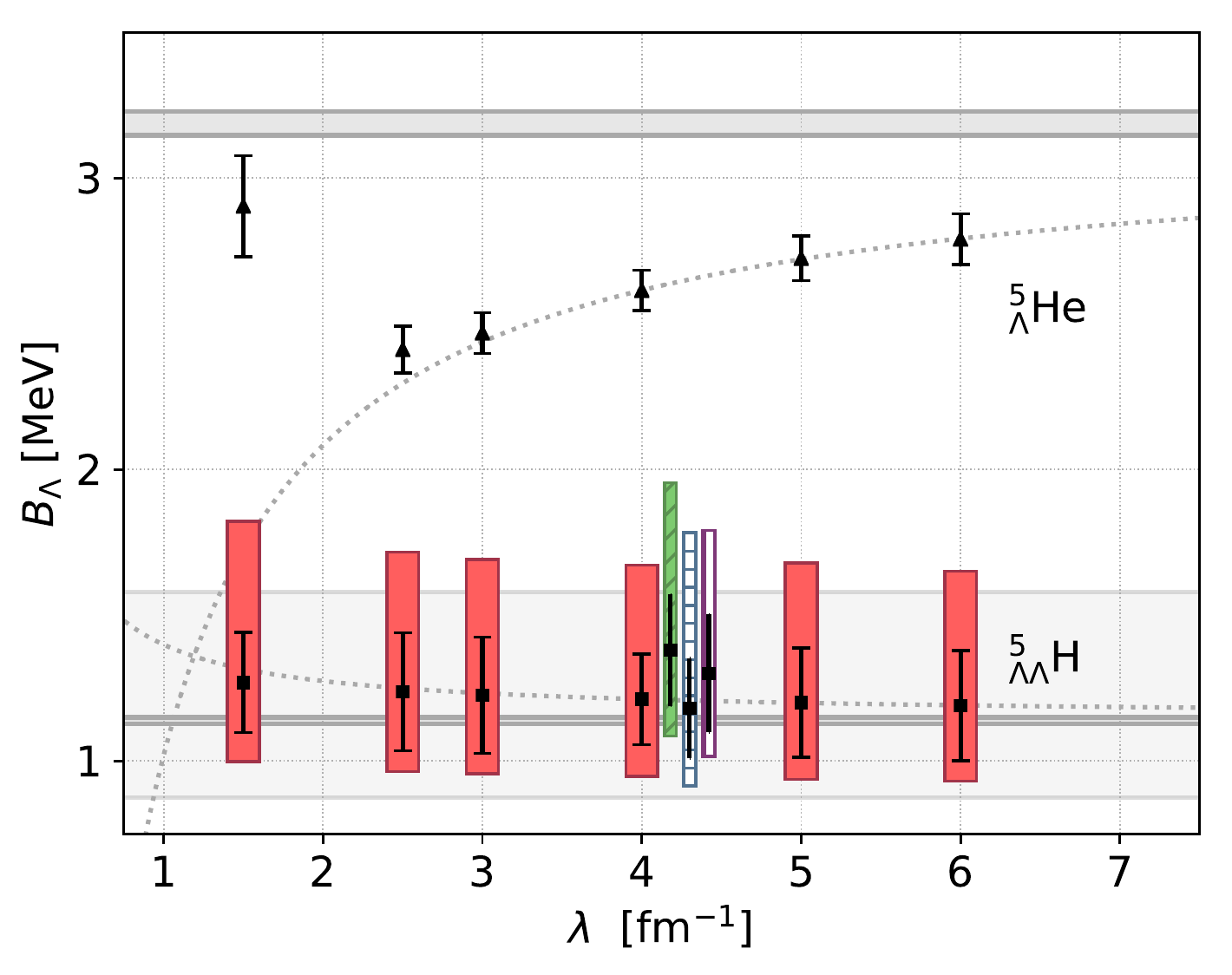}  
\caption{$\Lambda$ separation energies $B_\Lambda(^{~~5}_{\Lambda\Lambda}$H) 
and $B_\Lambda(^5_\Lambda$He) from SVM calculations that use {\nopieft} 
LO two-body (\ref{eq:V2}) and three-body (\ref{eq:V3}) regularized contact 
interactions, constrained by requiring $\Delta B_{\Lambda\Lambda}(^{~~6}_{
\Lambda\Lambda}$He)=0.67$\pm$0.17~MeV, are plotted as a function of the cutoff 
$\lambda$. Error bars (in black) reflect the experimental uncertainty inherent 
in the $^3_\Lambda$H, $^4_\Lambda$H, $^4_\Lambda$H$^\ast$ and $^{~~6}_{\Lambda
\Lambda}$He binding-energy input data, and (red) rectangles include also 
varying $a_{\Lambda\Lambda}$ between $-$0.5 to $-$1.9~fm. The $\Lambda N$ 
interaction model used is Alexander[B]~\cite{CBG18}, with results for models 
$\chi$LO, $\chi$NLO and NSC97f shown from left to right in this order for 
$\lambda$=4~fm$^{-1}$. Dotted lines show extrapolations, as $\lambda\to
\infty$, to the respective scale renormalization invariance limits marked 
by gray horizontal bands. The wider $^{~~5}_{\Lambda\Lambda}$H band 
accounts for uncertainties in the experimental values of binding energies 
used in extrapolating to $\lambda\to\infty$.} 
\label{fig:errors} 
\end{center} 
\end{figure}

Calculated values of the $\Lambda$ separation energy 
$B_\Lambda(^{~~5}_{\Lambda\Lambda}$H) are shown in Fig.~\ref{fig:errors}. 
Several representative values of the $\Lambda\Lambda$ scattering length were 
used: $a_{\Lambda\Lambda}$=$-0.5,-0.8,-1.9$~fm, spanning a broad range of 
values suggested in $\Lambda\Lambda$ correlation studies~\cite{GHH12,MFO15} 
of experimental spectra mentioned in the Introduction. Again, the choice of 
$a_{\Lambda\Lambda}$ determines the one $\Lambda\Lambda$ LEC required at LO, 
while the $\Lambda\Lambda N$ LEC was fitted to the $\Delta B_{\Lambda\Lambda}
(^{~~6}_{\Lambda\Lambda}$He)=0.67$\pm$0.17~MeV datum. 
For the $\Lambda N$ interaction terms we generally used the Alexander[B] 
$\Lambda N$ model~\cite{CBG18}, with its $\Lambda N$ $C_3$ and $C_4$ LECs 
(see Fig.~\ref{fig:diagrams}) corresponding to the scattering lengths 
$a_s$=$-1.8$~fm and $a_t$=$-1.6$~fm, respectively, from~\cite{Alex68}. 
For cutoff $\lambda$=4~fm$^{-1}$ we also used three other $\Lambda N$ 
interaction models from Ref.~\cite{CBG18}: $\chi$LO~\cite{PHM06}, 
$\chi$NLO~\cite{HPKMN13} and NSC97f~\cite{RSY99}, demonstrating that the 
$\Lambda N$ model dependence is rather weak when it comes to $\Lambda$ 
separation energies in double-$\Lambda$ hypernuclei, provided $B_\Lambda$ 
values of single-$\Lambda$ hypernuclei for $A<5$ are fitted to generate the 
necessary $\Lambda NN$ LECs.{\footnote{This is reminiscent of the weak 
dependence of $\Lambda$ separation energies in single-$\Lambda$ hypernuclei 
on the input $NN$ interaction, found in few-body calculations by Nogga et 
al.~\cite{Nogga02}.}} Calculated values of $B_\Lambda(^5_\Lambda$He), 
compatible with those from Ref.~\cite{CBG18} are also shown in the figure, 
demonstrating the suitability of the input $\Lambda N$ model. One observes 
that $^{~~5}_{\Lambda\Lambda}$H comes out particle stable over a broad range 
of finite cutoff values used in the calculations. This is not the case for 
$^{~~4}_{\Lambda\Lambda}$H which, as discussed above, is unbound with respect 
to $^3_\Lambda$H for most of the permissible parameter space. 

The calculated $B_\Lambda$ values shown in Fig.~\ref{fig:errors} exhibit 
renormalization scale invariance in the limit of $\lambda\to\infty$. 
To figure out the associated $B_\Lambda(\lambda\to\infty)$ values, 
we extrapolated $B_\Lambda(\lambda)$ for $\lambda \geq 4$~fm$^{-1}$ 
using a power series in the small parameter $Q/\lambda$:
\begin{equation} 
\frac{B_\Lambda(\lambda)}{B_\Lambda(\infty)}=\left[1+\alpha\frac{Q}{\lambda}
+\beta\left(\frac{Q}{\lambda}\right)^2+\ldots\right].  
\label{eq:extrap} 
\end{equation} 
The corresponding extrapolation curves are shown by dotted lines in 
Fig.~\ref{fig:errors}, converging at asymptotic values $B_\Lambda(\infty)$ 
given with their extrapolated uncertainties by the gray horizontal bands 
in the figure. $^{~~5}_{\Lambda\Lambda}$H remains particle stable in 
this limit with $\Lambda$ separation energy $B_\Lambda(\infty)=1.14\pm 
0.01^{+0.44}_{-0.26}$~MeV, where the first uncertainty is due to 
extrapolating by use of Eq.~(\ref{eq:extrap}) and the second one 
is due to the $a_{\Lambda\Lambda}$ and $B_{\Lambda}$ uncertainties. 

\begin{figure}[htb] 
\begin{center} 
\includegraphics[width=0.7\textwidth]{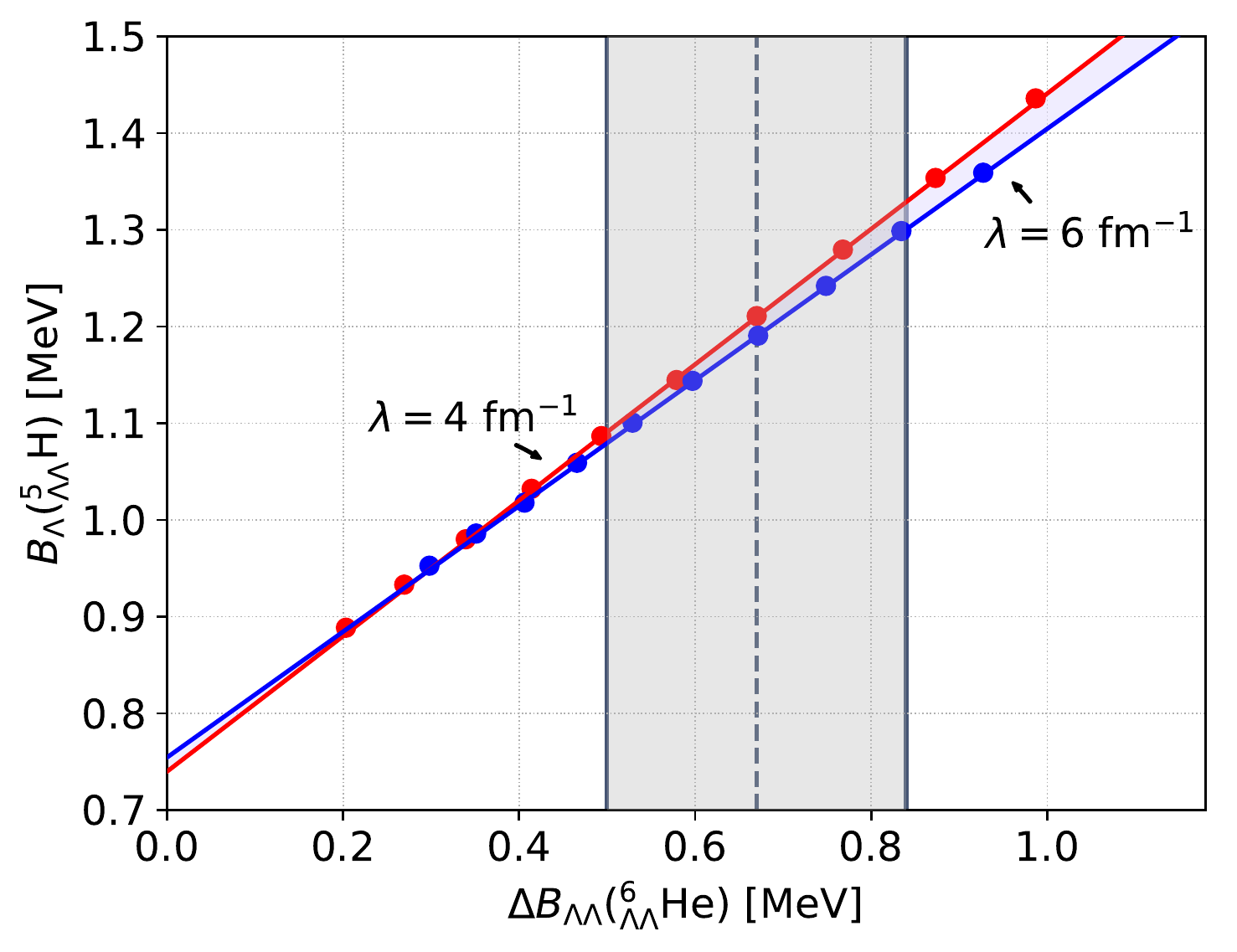} 
\caption{Hypernuclear Tjon lines: calculated $\Lambda$ separation energies 
$B_\Lambda(^{~~5}_{\Lambda\Lambda}$H) are plotted as a function of the 
constrained value assumed for $\Delta B_{\Lambda\Lambda}(^{~~6}_{\Lambda
\Lambda}$He) for two cutoff values, using $a_{\Lambda\Lambda}$=$-$0.8~fm. 
The shaded vertical area marks the observed value $\Delta B_{\Lambda\Lambda}
(^{~~6}_{\Lambda\Lambda}$He)=0.67$\pm$0.17~MeV. The $\Lambda N$ interaction 
model used is Alexander[B]~\cite{CBG18}.} 
\label{fig:Tjon} 
\end{center} 
\end{figure} 

The $\Lambda$ separation energies $B_\Lambda(^{~~5}_{\Lambda\Lambda}$H) 
studied above are correlated with those of $^{~~6}_{\Lambda\Lambda}$He 
in a way reminiscent of the Tjon line correlation between binding 
energies calculated for $^3$H and $^4$He~\cite{Tjon75}. This is shown 
in Fig.~\ref{fig:Tjon} by the linear dependence of $B_\Lambda(^{~~5}_{
\Lambda\Lambda}$H), for two given values of the cutoff $\lambda$, on the value 
assumed for $\Delta B_{\Lambda\Lambda}(^{~~6}_{\Lambda\Lambda}$He), which 
was varied for this purpose around the `physical' value 0.67$\pm$0.17~MeV. 
We note that the cutoff dependence of this correlation is very weak. 
The hypernuclear correlation noted here is generated by variation of the 
$\Lambda\Lambda N$ LEC which is derived from $\Delta B_{\Lambda\Lambda}
(^{~~6}_{\Lambda\Lambda}$He). This is similar to the origin and realization of 
Tjon-line correlations in nuclear physics, where many-body contact interaction 
terms beyond three-body terms do not appear at LO~\cite{PHM05}. However unlike 
other physics applications where Tjon lines were shown to hold, its appearance 
here does not require proximity to the unitary limit. 

We note that $a_{\Lambda\Lambda}$ includes implicitly the coupling of the 
$\Lambda\Lambda$ channel to the higher mass $I$=$S$=0 $\Xi N$ and $\Sigma
\Sigma$ channels. However, beginning with $^{~~6}_{\Lambda\Lambda}$He the 
coupling to the relatively low-lying $\Xi N$ channel is partially Pauli 
blocked (with the formed nucleon excluded from the $s$ shell). It could be 
argued then that the reference value of $\Delta B_{\Lambda\Lambda}(^{~~6}_{
\Lambda\Lambda}$He) used in this work has to be somewhat increased in order 
to account for the blocked states which are included effectively in the 
present LO application of {\nopieft} to $\Lambda\Lambda$ hypernuclei. 
The coupled-channel calculations by Vida\~{n}a et al.~\cite{VRP04} suggest 
an increase of $\approx$0.25~MeV which according to Fig.~\ref{fig:Tjon} 
would increase $B_\Lambda(^{~~5}_{\Lambda\Lambda}$H) by roughly 0.15~MeV 
and $^{~~4}_{\Lambda\Lambda}$H, had it been bound, by no more than 0.03~MeV. 
Interestingly, recent HAL QCD Collaboration studies based on LQCD find 
a vanishingly small $I=0$ $^1S_0$ $\Lambda\Lambda - \Xi N$ coupling potential 
for $r\gtrsim 0.6$~fm~\cite{Sasaki18}. Consequences of this exremely weak 
coupling, as well as the one derived at NLO within $\chi$EFT~\cite{HMP16}, 
on few-body $\Lambda\Lambda$ hypernuclei should be studied in future NLO 
calculations.

\section{Summary and outlook} 
\label{sec:sum} 

The focus in this first comprehensive $\nopieft$ application to light $\Lambda
\Lambda$ hypernuclei was to study the onset of binding in the ${\cal S}=-2$ 
hadronic sector by constraining $\Delta B_{\Lambda\Lambda}(^{~~6}_{\Lambda
\Lambda}$He) to the most recent value 0.67$\pm$0.17~MeV~\cite{HN18} assigned 
to the Nagara event~\cite{LL6He}. We varied the value assumed for the 
$\Lambda\Lambda$ scattering length $a_{\Lambda\Lambda}$ over a range of 
values suggested in Refs.~\cite{GHH12,MFO15} by analyzing $\Lambda\Lambda$ 
correlations observed in several high-energy production reactions, but barring 
an unlikely $\Lambda\Lambda$ bound state~\cite{halqcd12,HM12,Gal13}. 
Our results suggest with little model dependence that both members of the 
$A=5$ isodoublet pair, $^{~~5}_{\Lambda\Lambda}$H and $^{~~5}_{\Lambda\Lambda}
$He, are particle stable. Of the $A=4$ $\Lambda\Lambda$ hypernuclei, the 
particle stability of the $I=0$ $^{~~4}_{\Lambda\Lambda}$H(1$^+$) requires 
values of $|a_{\Lambda\Lambda}|\gtrsim 1.5$~fm, which are unlikely in our 
opinion. The $I=1$ excited state $^{~~4}_{\Lambda\Lambda}$H(0$^+$), or its 
isospin analog state $^{~~4}_{\Lambda\Lambda}$n are far from being bound; 
if any of these were established experimentally, the soundness of the Nagara 
event would suffer a serious setback. 

Extensions of the present LO work should consider explicit 
$\Lambda\Lambda$-$\Xi N$-$\Sigma\Sigma$ coupling in the $^1S_0$ channel or, 
at least, address momentum dependent $\Lambda\Lambda$ interaction components 
generated in NLO EFT through effective-range ($r_{\Lambda\Lambda}$) 
contributions. We note that no conclusive determination of $r_{\Lambda\Lambda}
$ exists yet because of the scarce and inaccurate hyperon-hyperon (mostly 
$\Xi^- p$) scattering and reaction data available in the $\approx$25~MeV 
interval between the $\Lambda\Lambda$ and $\Xi N$ thresholds. For example, 
small values of $r_{\Lambda\Lambda}$ between 0.3 to 0.8~fm were derived from 
such data in the LO $\chi$EFT work of the J\"{u}lich-Bonn group~\cite{PHM07} 
using values of $a_{\Lambda\Lambda}$ about $-$1.5~fm. In contrast, 
large values of $r_{\Lambda\Lambda}$ between 5 to 7~fm were derived from 
the same data in the NLO $\chi$EFT work of the J\"{u}lich-Bonn-Munich 
group~\cite{HMP16} using values of $a_{\Lambda\Lambda}$ about $-$0.65~fm. 
This dichotomy is apparent also for the Nijmegen soft core potentials 
listed in Table I of Ref.~\cite{GHH12} and would have to be considered 
in future studies of $\Lambda\Lambda$ hypernuclei.

\section*{Acknowledgments} 
The work of L.C. and N.B. was supported by the Pazy Foundation and by the 
Israel Science Foundation Grant No. 1308/16. The work of M.S. and J.M. was 
supported by the Czech Science Foundation GACR Grant No. 19-19640S.

\end{document}